\documentclass{article}
\usepackage{ijcai26}

\usepackage{times}
\usepackage{soul}
\usepackage{url}
\usepackage[hidelinks]{hyperref}
\usepackage[utf8]{inputenc}
\usepackage[small]{caption}
\usepackage{graphicx}
\usepackage{amsmath}
\usepackage{amsthm}
\usepackage{booktabs}
\usepackage{algorithm}
\usepackage{algorithmic}
\usepackage[table]{xcolor}
\usepackage{array} 
\usepackage{subcaption}
\usepackage{float}
\usepackage{placeins} % in the preamble
\usepackage[switch]{lineno}

\definecolor{shannoncol}{RGB}{210,230,255}   % light blue
\definecolor{mincol}{RGB}{255,220,220}       % light red
\definecolor{renyicol}{RGB}{220,255,220}     % light green

\newcommand{\SubFigVRule}[3]{%
  \hspace{#1}%
  \raisebox{#2}{\vrule width 0.8pt height #3 depth 0pt}%
  \hspace{#1}%
}

\makeatletter
\newcommand*\mysize{%
  \@setfontsize\mysize{8.0}{9.0}%
}
\makeatother

\title{When Numbers Start Talking: Implicit Numerical Coordination Among LLM-Based Agents}
\vspace{-2mm}
\author{
Alessio Buscemi$^1$, 
Daniele Proverbio$^2$, 
Alessandro Di Stefano$^3$, 
The Anh Han$^3$, 
German Castignani$^1$,
Pietro Liò$^4$\\
\affiliations
$^1$ Luxembourg Institute of Science and Technology, 
$^2$ University of Trento, 
$^3$ Teesside University, \\
$^4$ University of Cambridge\\
alessio.buscemi@list.lu,
daniele.proverbio@unitn.it,
a.distefano@tees.ac.uk,\\
t.han@tees.ac.uk,
german.castignani@list.lu,
pl219@cam.ac.uk
}
\date{}
\begin{document}

\maketitle

\begin{abstract}
Large Language Model (LLM)-based agents increasingly operate in multi-agent systems (MAS) where strategic interaction and coordination are required. While existing work has largely focused on individual agents or on interacting agents sharing explicit communication, less is known about how interacting agents coordinate implicitly. In particular, agents may engage in \textit{covert communication}, relying on indirect or non-linguistic signals embedded in their actions.
This paper presents a game-theoretic study of covert communication in LLM-driven MAS. We analyse interactions across four canonical game-theoretic settings under different communication channels, including explicit, restricted, and absent communication. Considering heterogeneous agent personalities and both one-shot and repeated games, we characterise when covert signals emerge and how they shape coordination and strategic outcomes.
\end{abstract}

\section{Introduction}
AI agents based on LLMs are increasingly deployed across academic \cite{lu2024llms}, social \cite{tessler2024ai}, and industrial domains \cite{hettiarachchi2025exploring,hughes2025ai}. This rapid adoption calls for robust frameworks capable of anticipating agent behaviour, both in interactions with humans and in environments involving multiple autonomous agents. Reliable behavioural prediction is a prerequisite for enabling new applications, supporting the deployment of trustworthy AI systems, and mitigating risks such as bias amplification or strategically undesirable outcomes \cite{fulgu2024surprising,andras2018trusting}.
Most existing efforts have focused on individual agents, with significant progress in transparency and interpretability \cite{ali2025entropy,el2025towards}, as well as in the detection of inconsistencies, biases, and hallucinations \cite{huang2025survey,zhou2024unibias}. In contrast, scenarios involving multiple interacting agents remain comparatively underexplored \cite{hammond2025multi}. In such settings, collective dynamics can give rise to emergent behaviours and biases that are not directly predictable from the properties of isolated agents. These interaction-driven phenomena may in turn lead to unreliable outcomes, including unfair decisions, systematic advantages for specific actors, or distortions of competitive equilibria. Methods based on the emulation of human behaviour \cite{park2023generative}, while informative, may therefore yield unreliable conclusions when applied to autonomous agents with distinct optimisation dynamics.
The relevance of these challenges is particularly evident in applied domains such as automated dispute resolution \cite{brooks2022artificial,falcao2024making}, auction design and pricing mechanisms \cite{bahtizin2019using,chen2023utility}, and negotiation processes in supply chains \cite{abaku2024theoretical,min2010artificial,ramachandran2022contract}.

Game theory \cite{owen2013game} provides a principled framework for modelling strategic multi-agent interactions and predicting the responses of rational agents \cite{balabanova2025media,falcao2024making,HanAICom2022emergent}. While traditionally applied to human decision-making \cite{stewart2024distorting,talajic2024strategic}, game-theoretic models have increasingly been adopted to study LLM-based agents in both classical games and more complex strategic environments \cite{fontana2024nicer,willis2025will}, including distributed and game-like computational systems \cite{he2025generative}.
Within this framework, it is well established that communication can fundamentally alter strategic behaviour compared to silent games \cite{farrell1996cheap,skyrms2010signals,song2026network}. Consistently, studies on AI-based games show that explicit communication plays a central role in enabling coordination and enhancing cooperation among LLM agents, relative to settings in which communication is restricted or absent \cite{sun2025game,buscemi2025strategiccommunicationlanguagebias}. These findings underscore the importance of information exchange in shaping collective behaviour and strategic efficiency in multi-agent systems.
Recent work has formalised the dynamics of LLM-based agents using analytical and game-theoretic models \cite{sun2025game}. While such approaches yield valuable theoretical insights, they necessarily abstract away from the full complexity of LLM behaviour and communication in order to remain tractable. At the opposite end of the spectrum, other studies have explored collaboration through high-capacity communication mechanisms, such as direct information exchange in continuous latent spaces, explicitly removing linguistic and symbolic constraints to maximise expressiveness and efficiency \cite{zou2025latent}.

In contrast, we focus on the complementary setting in which communication is severely constrained. Rather than simplifying agent behaviour or expanding communication capacity, we study the full complexity of LLM agents operating under minimal signalling affordances. Beyond explicit messaging, agents may also share information through more subtle and less transparent channels. We refer to this phenomenon as \emph{covert communication}, in which agents do not exchange information in human language but instead rely on indirect, non-linguistic, or structured signals. Such signals may function as implicit codes, enabling coordination without overt communication and potentially bypassing communication constraints or monitoring mechanisms. Covert communication is related to the concept of \textit{subliminal learning} \cite{cloud2025subliminal}, but focuses on coordination emerging from interaction rather than mutual influence, and extends prior studies on secret collusion among agents \cite{motwani2024secret}.
In this paper, we provide a systematic study of covert communication in MAS driven by LLMs. We investigate whether and how agents form implicit signalling strategies across four canonical game-theoretic settings spanning the standard spectrum of cooperation–defection dilemmas. We analyse agent behaviour under multiple communication regimes, including explicit communication, restricted numerical signalling, random numerical outputs, and communication-free baselines. To assess robustness, we further consider heterogeneous personality assignments and both one-shot and repeated interactions. This experimental design allows us to characterise when covert communication arises, how it interacts with incentive structures, and how it affects coordination and strategic outcomes under realistic communication constraints.

%\section{Background} % For the sake of space, it can be skipped, if we provide sufficient references in the introduction

\section{Methodology}
This section describes the methodology adopted in this study.

\subsection{Games Selection}
To provide a comprehensive range of strategic interactions under different incentive structures, we consider four canonical pairwise games: the \textit{Prisoner's Dilemma} (PD), \textit{Snowdrift} (SD), \textit{Stag Hunt} (SH), and \textit{Harmony} (H). Each game corresponds to a distinct ordering of incentives between cooperation and defection, leading to qualitatively different strategic regimes: strictly dominant defection (PD), mixed and anti-coordination incentives (SD), coordination under risk (SH), and strictly dominant cooperation (H) \cite{wang2015universal,HAN202633}. Together, these games are recognised in the game-theoretic literature to form a minimal and complete taxonomy of two-player social dilemmas, providing a standard reference framework for comparing strategic behaviour across fundamentally different interaction contexts.
Following common practice in the literature \cite{wang2015universal,sigmund2010calculus}, we adopt standard payoff configurations for each game, which preserve their canonical incentive structures while enabling a consistent comparison across settings. Referring to the payoff matrix entries, the Prisoner's Dilemma's penalties are defined as
\( x_{1,1} = (1, 1) \), \( x_{1,2} = (5, 0) \), \( x_{2,1} = (0, 5) \), and \( x_{2,2} = (3, 3) \).  
For the Snowdrift game, we use  
\( x_{1,1} = (3, 3) \), \( x_{1,2} = (0, 5) \), \( x_{2,1} = (5, 0) \), and \( x_{2,2} = (1, 1) \).  
For the Stag Hunt, penalties are set to  
\( x_{1,1} = (4, 4) \), \( x_{1,2} = (0, 3) \), \( x_{2,1} = (3, 0) \), and \( x_{2,2} = (2, 2) \).  
For the Harmony game, we consider  
\( x_{1,1} = (5, 5) \), \( x_{1,2} = (2, 3) \), \( x_{2,1} = (3, 2) \), and \( x_{2,2} = (1, 1) \).

\subsection{LLM Selection}
To enable direct interpretation of covert communication effects, we use a baseline model whose overt behaviour is already consistent with classical game-theoretic expectations. Without such a baseline, deviations induced by covert signalling cannot be reliably distinguished from inconsistent or non-strategic behaviours associated with the LLM itself. Following results from \cite{buscemi2025fairgame}, we use GPT-4o (version gpt-4o-2024-11-20) as the LLM that most closely adheres to expected game-theoretic behaviours across standard games, exhibiting greater stability and interpretability than alternative models.
Preliminary experiments on other LLMs (Claude, Mistral) across all four games were also conducted; however, we observed similar issues identified in \cite{buscemi2025fairgame}, including unstable strategies and systematic deviations from equilibrium behaviour. While interesting in their own right, these behaviours introduce confounds that are incompatible with the goals of the present study.
A model from the GPT family is also analysed with respect to Subliminal Learning \cite{cloud2025subliminal}, providing methodological continuity with prior work on implicit and hidden communication channels.

\subsection{Types of Communication}

\begin{table*}[h!]
\centering
\mysize
\setlength{\tabcolsep}{4pt}
\caption{Communication channels considered in the experiments.}
\begin{tabular}{p{1cm} p{3cm} p{5cm} p{8cm}}
\hline
\textbf{ID} & \textbf{Communication} & \textbf{Message source} & \textbf{Constraints / intent} \\
\hline

None & No communication 
   & None 
   & No messages exchanged between agents \\
\hline
NL & Natural language (EN) 
   & LLM-generated 
   & Explicit human language communication \\
\hline

C (D) & Covert (Dec) 
   & LLM-generated 
   & Explicit instruction to exchange numbers; no predefined semantics \\
\hline

C (H) & Covert (Hex) 
   & LLM-generated 
   & Explicit instruction to exchange numbers; no predefined semantics \\
\hline

LR (D) & Random output (Dec) 
   & LLM-generated 
   & Numbers generated without any communicative instruction \\
\hline

LR (H) & Random output (Hex) 
   & LLM-generated 
   & Numbers generated without any communicative instruction \\
\hline

R (D) & Injected random (Dec) 
   & External pseudorandom generator 
   & Numbers injected from outside; no agent control or intent \\
\hline

R (H) & Injected random (Hex) 
   & External pseudorandom generator 
   & Numbers injected from outside; no agent control or intent \\
\hline
\end{tabular}
\label{tab:communication}
\vspace{-4mm}
\end{table*}

We consider four types of communication between agents, summarised in Table \ref{tab:communication}, designed to progressively isolate the role of meaning, structure, and randomness in interactions. Each type of communication is conducted in its isolated set of experiments.
Initially, agents either communicate in natural language (English) or do not communicate at all. In the natural language setting, agents exchange unrestricted textual messages, enabling explicit and semantically grounded coordination. In contrast, the no-communication setting prohibits any message exchange. These channels follow established experimental designs used in prior work and provide a reference point for explicit coordination and its absence.

Then, we introduce covert communication. In this setting, agents are explicitly instructed to communicate at each round by outputting a sequence of exactly ten numbers. No predefined interpretation, encoding scheme, or semantic grounding is provided. Any meaning associated with these numerical sequences must therefore emerge endogenously through interaction. We consider two variants, where numbers are expressed either in decimal or hexadecimal form.
To set an independent baseline, a distinct experimental round asks agents to output a sequence of ten numbers without assigning any communicative purpose to them. Both decimal and hexadecimal representations are used. This condition controls for the mere presence of numbers in the interaction, without incentivising their use as a communication channel.

Finally, we introduce an external random baseline, where numerical sequences are injected from outside the agents using pseudo-random numbers generated via the Python \texttt{random} library. These numbers are independent of the agents' internal reasoning and are provided in either decimal and hexadecimal formats. This setting allows us to disentangle effects caused by agent-generated signals from those arising purely from exposure to random numerical input.
By comparing these conditions, we can assess whether numerical sequences influence behaviour simply by being present in the interaction, or whether they acquire strategic meaning only when agents intentionally generate them for communication. We hypothesised that covert communication would exhibit dynamics that differ from both agent-generated random outputs and externally injected pseudo-random sequences. Such differences would suggest the presence of structured behaviour inconsistent with pure noise.

\subsection{Implementation}

FAIRGAME~\cite{buscemi2025fairgame}, a framework for systematic simulations of strategic interactions between LLM-based agents, was chosen to implement the experiments. It provides a game-theory grounded environment while supporting heterogeneous agent attributes, interaction protocols, and communication constraints. Game scenarios are specified through prompt-driven interfaces, allowing agent characteristics, payoff structures, and interaction rules to be defined in natural language.
Each experimental setup is defined by a \emph{Configuration File}, which encodes game parameters, payoff matrices, and agent-level attributes, e.g. personality traits~\cite{fan2024comp,he2024afspp,newsham2025inducing}; and a \emph{Prompt Template}, which specifies the instructions presented to agents at each round. Prompt templates are dynamically instantiated using configuration parameters and, for repeated interactions, include a structured representation of the interaction history.
Covert communication and LLM-generated random output are realised by modifying prompt instructions, without changes to the FAIRGAME codebase. By contrast, externally injected random communication cannot be achieved through prompt engineering alone; we therefore extended the framework to inject pseudo-random numerical sequences generated outside the agents, ensuring independence from their internal reasoning.

\subsection{Experimental setting}

Following the experimental setup introduced in \cite{buscemi2025fairgame}, we assign each agent one of two fixed personalities: cooperative (C) or selfish (S). This results in three possible personality pairings: (C, C), (C, S), and (S, S). Each agent is unaware of the personality of the other agent. All experiments are conducted independently for each personality combination and communication type. 
For one-shot games (i.e., single-round interactions), each combination of personality pairing and communication type is repeated $N=50$ times to support statistical robustness. Given the three personality pairings and all communication types considered, this results in a total of 1,200 runs per game in the one-shot setting.
We additionally study repeated games with a fixed horizon of 10 rounds. The experimental setup remains identical for the same games, except that the number of repetitions is reduced to $N=20$ per personality pairing and communication type, so as to balance inference cost and statistical reliability. In total, this yields 480 runs per game in the repeated setting, corresponding to 4,800 interaction rounds per game. 
In all experiments, agents are explicitly informed of the number of rounds prior to play. Mean cooperation levels are calculated as follows: for each game $g$, communication type $c$, personality combination $p$, round $t$ (with $t=1$ for one-shot games) and repetition $r$, we compute a cooperation counter for each agent, $\mathcal{C} = 1$ if an agent's action is cooperative, $\mathcal{C} = 0$ otherwise, following the game-specific definitions of cooperation. Mean cooperation is then given as $\bar{\mathcal{C}}_{g,p,t,c} = \frac{1}{2N} \cdot \sum_{r=1}^N \mathcal{C}_{g,p,t,c,r}$, where the factor 2 is used to mediate over agents in pairwise interactions.
%We quantify cooperation by encoding each agent's action as a binary variable, assigning 1 to the action corresponding to cooperation and 0 to the action corresponding to defection. The interpretation of cooperation and defection is game-specific and follows the standard definitions for each game. Cooperation is then measured as the mean of this binary variable across agents. In repeated games, this quantity is computed per round.

\section{Results}
We analyse the results of different communication types, and their influence on behaviours in one-shot and repeated games.

\subsection{Communication randomness and structure}

\begin{table*}[t]
\centering
\mysize
\caption{Normalised entropies for one-shot and repeated games. Each cell reports \textit{one-shot $|$ repeated}. S = Shannon, M = Min, R2 = Renyi-2.}
\label{tab:entropy_merged}
\setlength{\tabcolsep}{4pt}
\begin{tabular}{
l|
>{\columncolor{shannoncol}}c
>{\columncolor{mincol}}c
>{\columncolor{renyicol}}c|
>{\columncolor{shannoncol}}c
>{\columncolor{mincol}}c
>{\columncolor{renyicol}}c|
>{\columncolor{shannoncol}}c
>{\columncolor{mincol}}c
>{\columncolor{renyicol}}c|
>{\columncolor{shannoncol}}c
>{\columncolor{mincol}}c
>{\columncolor{renyicol}}c
}
\hline
 & \multicolumn{3}{c|}{\textbf{Harmony}}
 & \multicolumn{3}{c|}{\textbf{Snowdrift}}
 & \multicolumn{3}{c|}{\textbf{Stag Hunt}}
 & \multicolumn{3}{c}{\textbf{Prisoner's Dilemma}} \\
\cline{2-13}
 & S & M & R2
 & S & M & R2
 & S & M & R2
 & S & M & R2 \\
\hline
C (D)
& .449$|$.209 & .159$|$.041 & .290$|$.080
& .539$|$.624 & .364$|$.411 & .451$|$.512
& .497$|$.483 & .303$|$.261 & .415$|$.379
& .394$|$.429 & .240$|$.220 & .292$|$.339 \\
C (H)
& .840$|$.792 & .408$|$.304 & .673$|$.561
& .926$|$.857 & .561$|$.529 & .849$|$.743
& .916$|$.885 & .501$|$.579 & .810$|$.796
& .916$|$.876 & .541$|$.594 & .813$|$.780 \\
LR (D)
& .955$|$.948 & .669$|$.709 & .910$|$.914
& .959$|$.946 & .684$|$.728 & .921$|$.913
& .960$|$.951 & .693$|$.724 & .924$|$.923
& .960$|$.951 & .696$|$.747 & .923$|$.924 \\
LR (H)
& .962$|$.937 & .655$|$.721 & .922$|$.903
& .957$|$.936 & .642$|$.727 & .912$|$.903
& .961$|$.935 & .714$|$.740 & .928$|$.901
& .959$|$.950 & .631$|$.747 & .913$|$.923 \\
R (D)
& .980$|$.994 & .818$|$.888 & .964$|$.989
& .980$|$.994 & .806$|$.900 & .965$|$.988
& .981$|$.994 & .846$|$.900 & .965$|$.989
& .981$|$.994 & .848$|$.888 & .965$|$.989 \\
R (H)
& .976$|$.987 & .799$|$.855 & .957$|$.975
& .973$|$.988 & .799$|$.875 & .952$|$.976
& .978$|$.987 & .820$|$.875 & .959$|$.975
& .975$|$.987 & .799$|$.859 & .956$|$.975 \\
\hline
\end{tabular}
\vspace{-4mm}
\end{table*}

We measure message randomness using R\'enyi-2 (collision) entropy
$H_2 = -\log \sum_{i=1}^{m} p^2(x_i)$, Shannon entropy
$H = -\sum_{i=1}^{m} p(x_i)\log p(x_i)$, and min-entropy
$H_{\min} = -\log \max_i p(x_i)$, computed over the distribution of numerical messages $x_i$ under each communication type and game. Here, $x_i$ is one among all numerical messages exchanged under each communication type, separately for each game. All entropy measures are scaled by the maximum entropy $\log m$, attained under a uniform distribution over $m$ symbols, to lie in $[0,1]$.
For ${\text{R2, S, M}} \in[0,1]$, values closer to $1$ indicate higher entropy and therefore greater randomness, while values closer to $0$ indicate lower entropy and more structured, predictable sequences. For each game and communication type, entropies are computed by aggregating all messages produced by both agents across all runs, yielding 3,000 individual numbers per condition in the one-shot setting, and an equivalent aggregation across all rounds and repetitions in the repeated setting.

Table \ref{tab:entropy_merged} reports the results.
From a game-theoretic perspective, entropy here should be interpreted at the level of the communication channel rather than at the level of strategies or actions. Since we measure entropy over the distribution of numerical messages, it captures how strongly the \textit{symbolic output} is structured, rather than how agents mix over strategic choices. Lower entropy indicates that agents repeatedly reuse a restricted subset of numerical symbols, suggesting structured signalling induced by strategic interaction, while higher entropy corresponds to more diffuse use of the available symbol space, consistent with weak coordination pressure or noise in the message generation process \cite{fujimoto2024game}. 

From Table~\ref{tab:entropy_merged}, we observe that externally injected random numbers achieve entropy values very close to the theoretical maximum across all games, metrics, and communication types, in both one-shot and repeated play. This is expected, as these sequences are generated independently of the agents’ reasoning processes.
LLM-generated random outputs also exhibit high entropy in all settings, but consistently remain below the externally injected random baseline. This gap persists across games and entropy measures: while these sequences appear largely random, they nonetheless retain mild residual structure induced by the model’s generative process. Notably, this effect is stable across one-shot and repeated interactions, suggesting that repetition alone does not eliminate such residual regularity.
By contrast, covert communication exhibits markedly lower entropy than both random baselines. This effect is observed consistently across all four games and across all entropy measures, and is particularly pronounced in C(D). The reduction in entropy indicates systematic deviations from randomness, consistent with the use of structured signalling. When moving from one-shot to repeated games, entropy under covert communication generally decreases further or remains comparably low, suggesting that repeated interaction does not randomise the signals but instead preserves, and in some cases, reinforces their structure. C(H) exhibits higher entropy than C(D) in both settings, but remains clearly distinct from random baselines, indicating partial structure despite the larger symbols set.

Taken together, these results reveal a robust ordering of randomness that holds across games, entropy measures, and communication channels: injected random numbers exhibit the highest entropy, followed by LLM-generated random outputs, while covert communication consistently yields the lowest entropy. The stability of this ordering across one-shot and repeated games suggests that numerical sequences only acquire strong and persistent structure when agents are explicitly instructed to use them for communication. This observation motivates the subsequent analysis of whether such structured signals are associated with systematic changes in strategic behaviour and coordination.

\subsection{One-shot games}
\label{sub:result-os}

\begin{figure*}[t]
    \centering
    \begin{subfigure}[t]{0.46\textwidth}
        \centering
        \caption{One-shot games}
        \includegraphics[width=\linewidth]{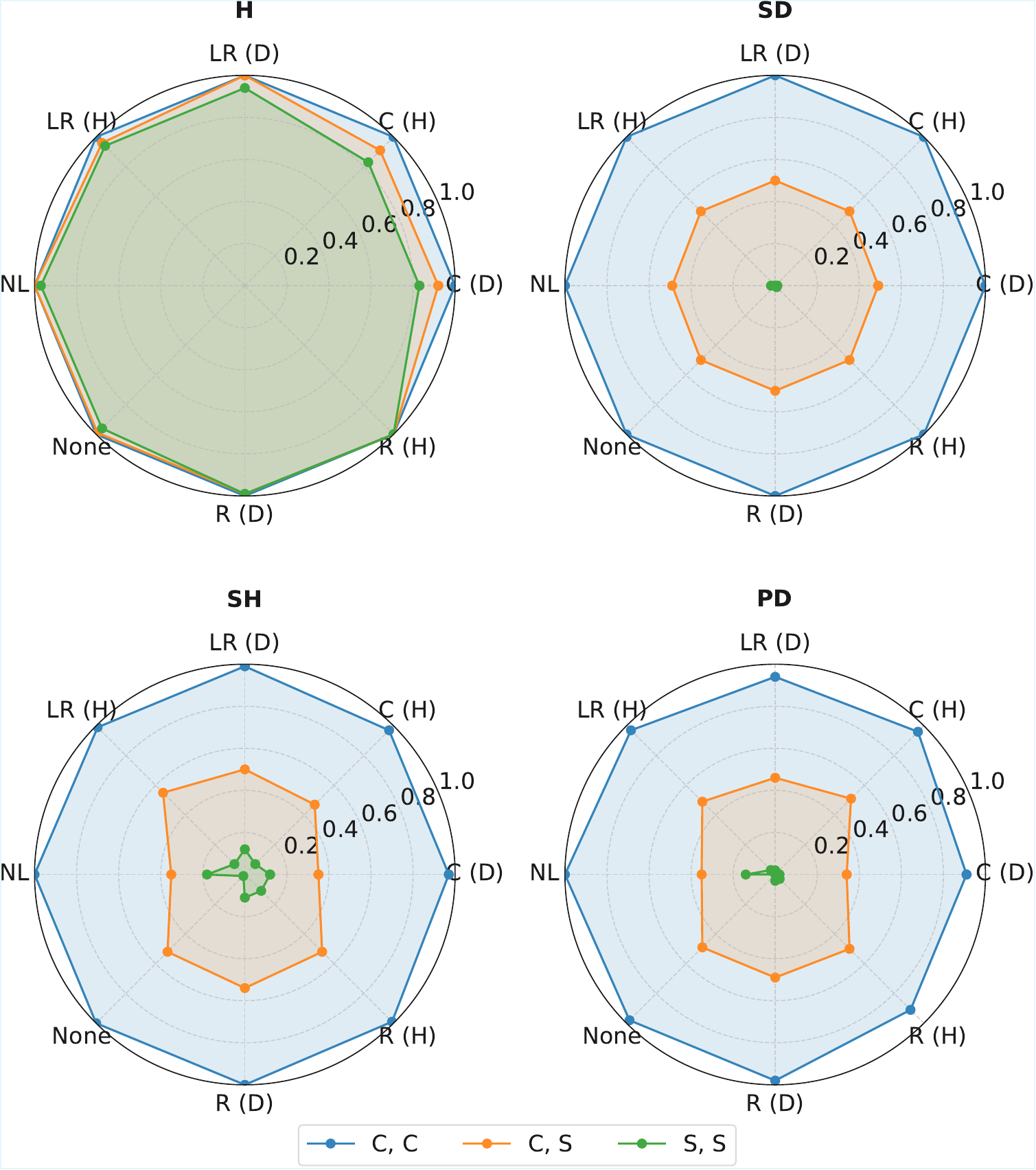}
        \label{fig:radarplots-oneshot}
    \end{subfigure}
    \hfill
    \unskip
    \SubFigVRule{0.8em}{-9.5cm}{9cm}
    \begin{subfigure}[t]{0.46\textwidth}
        \centering
        \caption{Repeated games}
        \includegraphics[width=\linewidth]{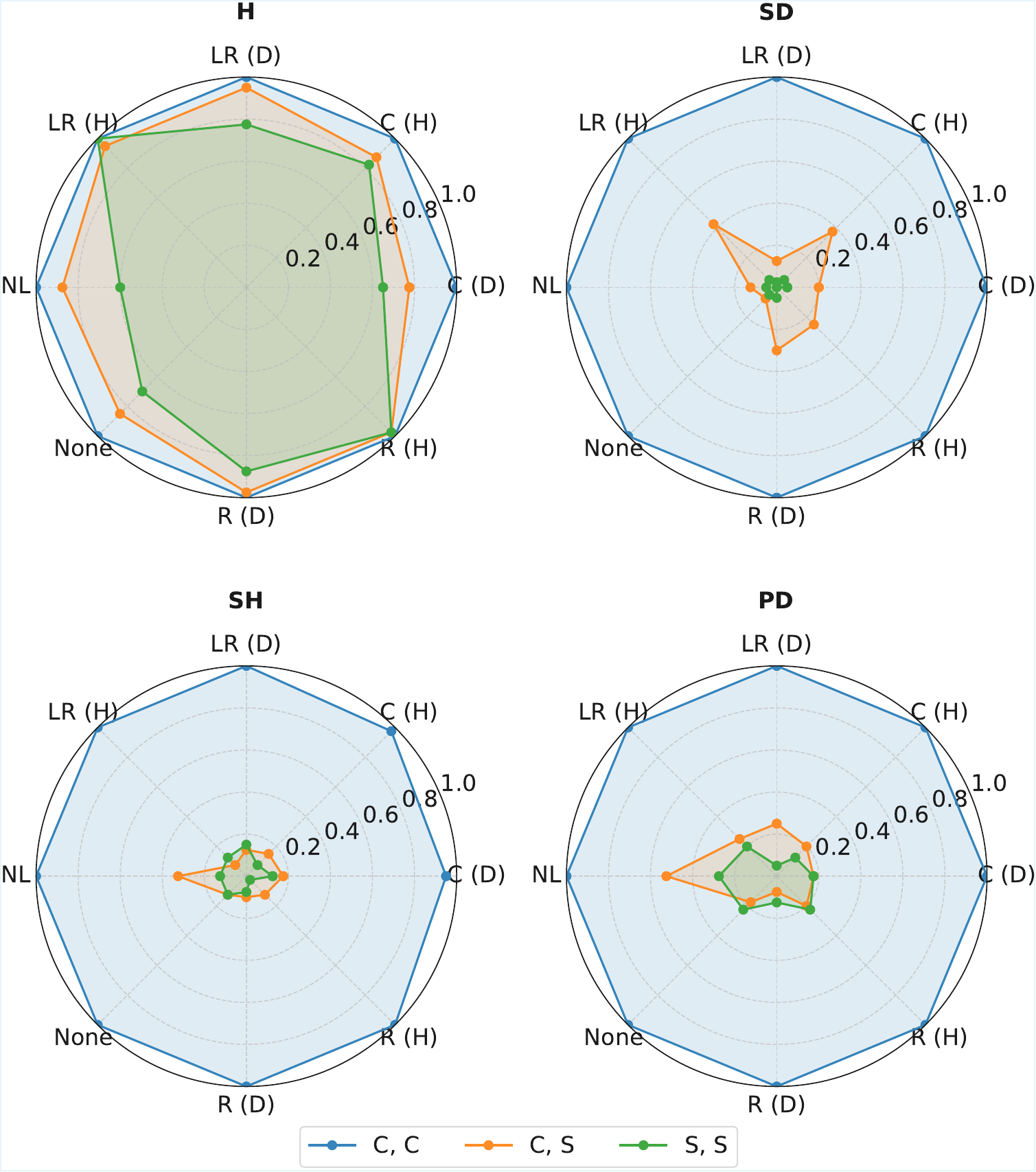}
        \label{fig:radarplots-repeated}
    \end{subfigure}
    \vspace{-4mm}
    \caption{Mean cooperation $\bar{\mathcal{C}}_{g,p,t,c}$ (pure Cooperation = 1, pure Defection = 0) by communication type $c$ (on the axes of the radar plot, see Tab.~\ref{tab:communication} for legend) and personality combinations $p$ (colour-coded) across the four games $g=\{$H, SD, SH, PD$\}$, for one-shot or repeated games -- respectively, in the (a) and (b) group. Each radar plot compares the impact of personalities along each communication type, while comparing plots in pairs shows the impact of game repetition (e.g., in SH game, the (C,S) personality, in orange, the radar is wider over all dimensions for one-shot rather than repeated games).}
    \label{fig:radarplots}
    \vspace{-4mm}
\end{figure*}

Fig. \ref{fig:radarplots-oneshot} reports the mean level of cooperation $\in [0,1]$ across communication types and personality combinations for the four games, computed by averaging the agents' decisions in one-shot scenarios.
In the Harmony game (H), communication types only marginally affects the expected strictly dominant cooperation, which remains high across all channels, with a slight reduction associated with the mixed and selfish personality scenarios when numerical sequences are generated by the agents, with slightly higher impact of covert communication than random output. Here, communication provides little impact and may even introduce slight noise.

In the Snowdrift game (SD), communication has no discernible impact on cooperation levels. The dominant factor is the personality combination, where including selfish players drastically reduces cooperation. Instead, within each personality pairing, cooperation remains remarkably stable across all communication types, and equilibrium behaviour in SD is largely insensitive to both explicit and implicit signalling.

The Stag Hunt (SH) game exhibits a more nuanced pattern. In the mixed personality scenario, cooperation is visibly reduced under natural language communication, as well as under covert communication with decimal numbers C(D), and to a lesser extent under covert hexadecimal communication C(H). Results for the selfish pairing are more heterogeneous, with no single communication channel consistently dominating. This suggests that having any type of structured communication (compared to none or random) may interfere with equilibrium selection in coordination games, particularly when incentives are misaligned.

Prisoner's Dilemma (PD) shows a pattern similar to Stag Hunt for mixed personalities: natural language and covert decimal communications yield a clear reduction in cooperation. Instead, for the selfish pairing, natural language communication increases cooperation (compared to other channels), while numerical or random communications exhibit no such effect. Hence, explicit linguistic communication can sometimes promote cooperation even in strictly dominant defection settings, whereas numerical signalling does not.
Compared to SH, in the PD game communication has a weaker impact on enhancing cooperation, which is in line with previous game-theoretical predictions \cite{hernandez2019cooperative}.  

Overall, these results indicate that communication affects behaviour primarily in games where coordination or strategic uncertainty is present, and that its influence depends not only on the channel but also on how agents interpret and exploit that channel within a given incentive structure. This limited impact of communication in one-shot interactions is consistent with game-theoretic results on cheap talk: in dominant-strategy games such as the PD, non-binding communication does not alter equilibrium outcomes, while in mixed-incentive games communication alone is insufficient to resolve strategic tension \cite{song2026network}.

\begin{table*}[t!]
\centering
\mysize
\caption{Top 5 most frequent symbols in one-shot games for covert communication (percentage of total)}
\renewcommand{\arraystretch}{1.05}
\setlength{\tabcolsep}{2pt}
\small
\begin{tabular*}{\textwidth}{@{\extracolsep{\fill}} lccccc ccccc}
\toprule
 & \multicolumn{5}{c}{\textbf{C (D)}} & \multicolumn{5}{c}{\textbf{C (H)}} \\
\cmidrule(lr){2-6} \cmidrule(lr){7-11}
\textbf{Game} & 1st & 2nd & 3rd & 4th & 5th & 1st & 2nd & 3rd & 4th & 5th \\
\midrule
\textbf{H} &
5 $|$ 65.1\% & 1 $|$ 14.2\% & 0 $|$ 8.83\% & 2 $|$ 4.83\% & 3 $|$ 3.77\% &
5A $|$ 6.83\% & 1A $|$ 3.84\% & 0A $|$ 2.92\% & 1F4 $|$ 2.38\% & 5 $|$ 2.35\% \\

\textbf{SD} &
3 $|$ 25.48\% & 0 $|$ 25.38\% & 5 $|$ 17.02\% & 1 $|$ 14.48\% & 2 $|$ 5.22\% &
1A3 $|$ 2.13\% & 1F4 $|$ 0.81\% & 3C $|$ 0.78\% & 1A $|$ 0.78\% & 3E8 $|$ 0.74\% \\

\textbf{SH} &
4 $|$ 32.7\% & 0 $|$ 22.13\% & 1 $|$ 20.43\% & 3 $|$ 12.17\% & 2 $|$ 6.02\% &
1A3 $|$ 3.3\% & 2B $|$ 1.35\% & 1A2 $|$ 1.35\% & 2B4 $|$ 1.04\% & 1F4 $|$ 1.04\% \\

\textbf{PD} &
1 $|$ 41.53\% & 0 $|$ 40.67\% & 2 $|$ 4.73\% & 5 $|$ 4.17\% & 3 $|$ 2.87\% &
1A3 $|$ 2.42\% & 3E8 $|$ 2.09\% & 0 $|$ 1.68\% & 1F4 $|$ 1.62\% & 1A2 $|$ 1.35\% \\
\bottomrule
\end{tabular*}
\normalsize
\label{tab:top5_covert_oneshot}
\vspace{-4mm}
\end{table*}

To better understand how agents exploit covert channels in this setting, Table~\ref{tab:top5_covert_oneshot} reports the five most frequent symbols generated under covert communication in the one-shot games, expressed as percentages of all transmitted messages. In the decimal variant, symbol distributions are strongly concentrated across all games, with one or two values accounting for a large fraction of messages. In H and PD, two dominant symbols together represent more than 80\% of all transmissions, while SD and SH exhibit slightly broader but still highly skewed distributions. This indicates that agents converge on compact, low-entropy signalling schemes even in single-shot interactions.
By contrast, the hexadecimal variant yields substantially flatter distributions. No single symbol exceeds a few percent of total messages, and the top five symbols account for only a limited share of the overall traffic. This suggests that agents exploit the larger representational space to distribute signals across a wider set of symbols, reducing concentration relative to C(D).
This suggests that, under decimal communication, agents tend to reuse values drawn from the payoff structure when constructing their signalling schemes, showing a statistical preference for payoff-related symbols. This link is broken when using hexadecimals, which are instead more randomly generated, but still retain some covert meaning. When interpretability is concerned, however, even with full knowledge of the payoff matrices the communicative logic remains far from transparent: it is unclear how these values are selected, combined, or interpreted within the agents' decision processes. The exchanges thus reveal a hybrid form of communication: somehow anchored in the environment but overall opaque to human reasoning. In this sense, interactions are \textit{partially covert}: not because humans lack access to the mapping, but because the functional role of these numbers in coordinating strategies does not follow human-intuitive patterns.

When considered jointly, the behavioural and symbolic results point to a coherent picture of covert communication in one-shot games. In H and SD, where cooperation is either dominant or largely insensitive to strategic uncertainty, communication has little effect on behaviours and correspondingly gives rise to either trivial or weakly structured signalling patterns. In these settings, concentrated decimal codes reflect stable action choices rather than meaningful coordination, while the flatter hexadecimal distributions indicate unused signalling capacity.
In contrast, in SH and PD, where equilibrium selection and strategic uncertainty are central, the behavioural sensitivity to communication is mirrored by more pronounced structure in covert signalling. The reduction in cooperation observed under natural language and covert decimal communication for mixed personalities aligns with the emergence of highly concentrated symbol usage, suggesting that agents form compact but opaque signalling conventions that influence strategic choices. Hexadecimal communication, while still covert, produces more diffuse symbol usage and correspondingly weaker behavioural effects, indicating that increased representational freedom does not necessarily translate into clearer or more effective coordination.

\subsection{Repeated games}

\begin{table*}[h]
\centering
\mysize
\caption{Top 5 most frequent symbols in repeated games for covert communication types (D or H), with their frequency over total messages.}
\renewcommand{\arraystretch}{1.05}
\setlength{\tabcolsep}{2pt} % minimal space between columns
\small
\begin{tabular*}{\textwidth}{@{\extracolsep{\fill}} lccccc ccccc}
\toprule
 & \multicolumn{5}{c}{\textbf{C (D)}} & \multicolumn{5}{c}{\textbf{C (H)}} \\
\cmidrule(lr){2-6} \cmidrule(lr){7-11}
\textbf{Game} & 1st & 2nd & 3rd & 4th & 5th & 1st & 2nd & 3rd & 4th & 5th \\
\midrule
\textbf{H} &
5 $|$ 83.4\% & 2 $|$ 5.6\% & 3 $|$ 1.9\% & 200 $|$ 1.2\% & 100 $|$ 1.1\% &
5A $|$ 12.5\% & 1A $|$ 3.4\% & 5A3 $|$ 1.6\% & 2A $|$ 1.6\% & 7A $|$ 1.6\% \\
\textbf{SD} &
3 $|$ 45.2\% & 5 $|$ 29.1\% & 0 $|$ 16.3\% & 1 $|$ 2.3\% & 2 $|$ 1.2\% &
3A $|$ 1.8\% & 2B $|$ 1.3\% & 8B $|$ 1.2\% & 3C $|$ 1.2\% & 1A $|$ 1.2\% \\
\textbf{SH} &
4 $|$ 36.3\% & 1 $|$ 18.1\% & 0 $|$ 13.4\% & 3 $|$ 13.1\% & 2 $|$ 9.3\% &
3A $|$ 1.7\% & 1A $|$ 1.6\% & 4A $|$ 1.6\% & 7A $|$ 1.5\% & 1F4 $|$ 1.5\% \\
\textbf{PD} &
0 $|$ 37.3\% & 1 $|$ 36.8\% & 5 $|$ 10.1\% & 2 $|$ 8.4\% & 3 $|$ 2.7\% &
1A $|$ 2.6\% & 2B $|$ 1.9\% & 7A $|$ 1.9\% & 8B $|$ 1.9\% & 3C $|$ 1.8\% \\
\bottomrule
\end{tabular*}
\normalsize
\label{tab:top5}
\vspace{-4mm}
\end{table*}

For repeated games, we compare the decisions made at the final round of each interaction. Fig.~\ref{fig:radarplots-repeated} reports the resulting mean cooperation levels per game and communication channel.
A first notable observation is that, while full cooperation is preserved for the (C,C) pairing across all games and communication types, cooperation for mixed and (S,S) pairings generally decreases compared to the one-shot setting. This indicates that repeated interaction does not necessarily promote cooperation and can, in many cases, discourage it. In the H game, natural language and covert communication are associated with the largest reductions in cooperation for the (S,S) pairing, and even the mixed (C,S) pairing is impacted.

For what concerns SD, communication has virtually no effect in the one-shot setting; by contrast, in repeated interactions, clear differences emerge across communication types. Natural language, LLM-generated random output, covert decimal communication, and no-communication are associated with substantially lower cooperation than other numerical communication variants, indicating that repetition amplifies sensitivity to how information is exchanged.
For the SH game, the mixed personality pairing has natural language communication leading to higher cooperation compared to all other communication types, which remain clustered at lower levels. For the (S,S) pairing, C(H) and R(H) yield the lowest cooperation levels, while the remaining channels exhibit similarly low outcomes.
A comparable structure is observed in PD. For (C,S), natural language communication again results in higher cooperation relative to other communication types. For the (S,S) pairing, LR(D), C(H), C(D), R(H) and R(D) are associated with the lowest cooperation levels, whereas natural language consistently yields higher cooperation.

Repeated-game results indicate that repetition does not systematically promote cooperation; rather, in most cases, it is associated with reduced cooperation compared to the one-shot setting. Differences across communication channels also become more pronounced under repetition, particularly for mixed personality pairings. Natural language communication remains the most distinctive channel, yielding relatively higher levels of cooperative behaviour, while numerical communication, whether covert or random, tends to produce similar and generally lower cooperation levels.

% \begin{table*}
% \centering
% \caption{Pearson correlation with the natural-language baseline by communication type.}
% \label{tab:pearson_comm}
% \footnotesize
% \begin{tabular}{lccccccc}
% \toprule
% \textbf{Comm. Type} 
% & \textbf{C (D)} 
% & \textbf{C (H) }
% &\textbf{ LR (D) }
% & \textbf{LR (H)} 
% & \textbf{R (D) }
% & \textbf{R (H) }
% & \textbf{None} \\
% \midrule
% \textbf{Pearson} 
% & 0.486 
% & 0.381 
% & 0.374 
% & 0.114 
% & 0.346 
% & 0.299 
% & 0.435 \\
% \bottomrule
% \end{tabular}
% \end{table*}

To get insights about the dynamics and co-evolution of cooperation under natural or numerical communication, we compute the Pearson correlation coefficients $\rho_{X,Y_j} \in [-1, 1]$ to summarise whether the series $X=\{x_1, \dots, x_{10}\}$ of cooperation levels $x_i (i=\{1, \dots, 10 \}$ rounds) under natural communication correlate with the series $Y_j=\{y_{1,j}, \dots, y_{10,j}\}$ of cooperation levels for each $j$-th numerical communication type.
%Table~\ref{tab:pearson_comm} reports Pearson correlations between strategy trajectories obtained under natural-language communication and those observed under alternative communication conditions. 
Correlations are computed \emph{within the same game and personality setting}, ensuring that comparisons are made between series generated under identical incentive structures and agent profiles. We exclude the (C,C) personality setting, as it typically converges immediately to full cooperation across games and communication types; this would trivially inflate similarity measures and obscure differences in coordination dynamics. %We use Pearson correlation because all trajectories are defined on the same decision scale, and the metric captures similarity in temporal dynamics and co-movement rather than absolute level differences, making it well suited for this setting.
The results are: $\rho_{X,C(D)}=0.486$, $\rho_{X,C(H)}=0.381$, $\rho_{X,LR(D)}=0.374$, $\rho_{X,LR(H)}=0.114$, $\rho_{X,R(D)}=0.346$, $\rho_{X,R(H)}=0.299$, $\rho_{X,\text{None}}=0.435$.

Overall, correlation values remain moderate, indicating that none of the channels fully reproduces the dynamics observed under natural-language interaction. Among the considered settings, covert communication based on agent-generated numerical sequences yields the highest correlation with the natural-language baseline, with the decimal variant achieving the strongest alignment. While this value is not high in absolute terms, it is consistently larger than those observed for random or non-intentional numerical outputs, suggesting that structured numerical exchanges can partially support coordination when agents are explicitly instructed to use them for communication.
Notably, also no-communication exhibits relatively high correlation. This indicates that a substantial portion of agent behaviour is driven by the underlying game structure and strategic incentives rather than by communication alone. Hence, improvements attributable to communication, whether explicit or covert, should be interpreted relative to this baseline rather than in isolation.

Conditions involving numerical output without communicative intent and externally injected pseudo-random sequences show lower correlations, reinforcing the view that numerical signals acquire strategic relevance only when agents intentionally generate them for coordination. Taken together, these results suggest that covert communication contributes to alignment with natural-language behaviour, but its effect remains limited and context-dependent. 
To better understand how agents exploit covert channels in repeated games,
Table \ref{tab:top5} reports the five most frequent numerical messages produced under covert communication in the repeated‑game setting, expressed as percentages of all transmitted messages. %In the decimal communication, distributions are relatively concentrated, typically one or two values account for one‑third to one‑half of all messages, and only in the Harmony game does a single number (5) dominate with more than 80\%. Hexadecimal communication, instead, yields flatter distributions with no symbol exceeding about 13\%, indicating greater dispersion.
Compared to the one-shot setting, repeated interactions amplify the structural differences observed between decimal and hexadecimal covert communication. In the decimal condition, symbols distributions in repeated games are more concentrated than in one-shot play. While one-shot interactions already exhibit skewed usage patterns, repetition leads to a stronger collapse onto a minimal set of symbols, most notably in the Harmony game, where a single value accounts for more than 80\% of all messages. This suggests that repeated exposure reinforces stable, low-entropy signalling conventions.
By contrast, hexadecimal communication remains comparatively dispersed under repetitions. No symbol becomes dominant, and only one top-ranked value ("5A" in the H game) exceed low double-digit percentages. This persistence of flatter distributions suggests that the larger symbol space inhibits convergence to compact codes even when agents interact repeatedly.

Taken together, these results indicate that repetition does not fundamentally alter the nature of covert communication, but instead sharpens pre-existing tendencies. Repeated games intensify the consolidation of signalling schemes in constrained numerical spaces, while richer representational channels preserve dispersion and limit dominance effects. As in the one-shot setting, the resulting codes remain semantically anchored in the payoff structure yet pragmatically opaque, indicating that repeated interaction strengthens coordination without necessarily improving interpretability.
This pattern is consistent with repeated-game scenarios in which agents exhibit stable, inferable traits, approximating near-complete-information play. In such condition, repetition primarily amplifies existing coordination dynamics rather than reshaping them through uncertainty about types or reputational incentives, highlighting a distinctive interaction between communication constraints and repeated play in LLM-based multi-agent systems \cite{willis2025will,fontana2024nicer}.

%\section{Discussion} % we already discuss the results before; we can directly go to the conclusions. Now we have half a column more space: we can extend the discussion a bit, strengthening the significance for interpretability and the interpretation in terms of game theory alignment @the ahn, Alessandro

\section{Conclusion}

This paper presents a systematic experimental analysis of numerical signalling in multi-agent interactions involving LLM-based agents under deliberately constrained communication. Across four canonical game-theoretic environments and multiple interaction conditions, we compare natural-language exchange, no communication, and numerical channels designed to disentangle intentional signalling from numerical noise. We show that numerical messages exhibit stable, non-random structure only when agents are explicitly instructed to use them for communication. In these cases, numerical outputs display consistently lower entropy than both LLM-generated random outputs and externally injected noise.

Three main observations emerge. First, representational constraints strongly influence the structure of numerical signalling. Decimal communication yields highly concentrated symbol distributions, often dominated by a small subset of values, whereas hexadecimal communication remains substantially more dispersed. Repeated interaction amplifies this contrast: decimal distributions become more concentrated, in some cases collapsing onto a single dominant symbol, while hexadecimal signalling preserves broader symbol usage.
Second, the behavioural impact of numerical communication is selective and context-dependent. In dominant-strategy games, communication has little effect on outcomes. In coordination-sensitive settings, however, differences across communication channels become visible, particularly for mixed personality pairings. In these cases, covert decimal communication produces behavioural patterns that are more closely aligned with those observed under natural-language communication than those arising from random or non-intentional numerical outputs. Nonetheless, correlations with natural-language dynamics remain moderate, indicating only partial alignment rather than functional equivalence.
Third, repeated interaction does not systematically improve cooperation. While repetition consolidates signalling patterns and reduces symbolic uncertainty, it is frequently associated with lower cooperation relative to one-shot interactions. This suggests that stabilised expectations do not necessarily translate into more efficient or socially desirable outcomes, and that repetition primarily amplifies existing behavioural tendencies rather than reshaping them.

Taken together, these findings indicate that numerical message sequences can acquire persistent structure and strategic relevance when agents intentionally use them for communication. However, the relationship between symbol usage and strategic intent remains opaque: even when numerical outputs are clearly non-random and correlated with behavioural change, their semantic interpretation is difficult to reconstruct, including when the payoff structure is fully known. As a result, behavioural regularities may be observable at an aggregate level while remaining resistant to post hoc semantic explanation.
From a broader perspective, these results highlight limits to the assumption that restricting natural-language communication alone is sufficient to constrain coordination in LLM-based MAS. At the same time, the effects of numerical signalling are neither uniform nor robust across settings, underscoring their dependence on incentive structure, representational constraints, and interaction regime. Future work should therefore examine how different architectural choices, training signals, or monitoring strategies influence the emergence and observability of such signalling in more deployment-relevant environments.

% \FloatBarrier

\bibliographystyle{named}
\bibliography{ijcai19}

\end{document}